\documentstyle[12pt]{article}

\newcommand{\be}{\begin{equation}}
\newcommand{\ee}{\end{equation}}
\makeatletter
\@addtoreset{equation}{section}
\makeatother
\def\thesection   {\arabic{section}}

\renewcommand{\theequation}{\thesection.\arabic{equation}}
\newcommand{\complex}{{\kern .1em {\raise .47ex
\hbox {$\scriptscriptstyle |$}}
    \kern -.4em {\rm C}}}
\newcommand{\real}{{{\rm I} \kern -.19em {\rm R}}}
\newcommand{\rational}{{\kern .1em {\raise .47ex
\hbox{$\scripscriptstyle |$}}
    \kern -.35em {\rm Q}}}
\renewcommand{\natural}{{\vrule height 1.6ex width .05em depth 0ex \kern -.35em
{\rm N}}}
\def\bea{\begin{eqnarray}}
\def\eea{\end{eqnarray}}
\def\bb#1{\hbox{\mybb#1}}
\def\complex{\bb{C}}

\def\real{\bb{R}}
\def\rational{\bb{Q}}
\def\R4{\bb{R}^4}
\def\unita{{1 \kern-.30em 1}}
\def\f{\phi}
\def\o{\omega}
\def\O{\Omega}
\def\p{\partial}
\def\a{\alpha}
\def\b{\beta}
\def\nn{\nonumber}
\def\th{\theta}
\def\s{\sigma}
\def\g{\gamma}
\def\G{\Gamma}
\def\l{\lambda}

\def\d{\delta}
\def\Si{\Sigma}
\begin{document}
%%%%%%%%%%%%%%%%%%%%%%%  FRONTESPIZIO %%%%%%%%%%%%%%%%%%%%%%%%%%%%%%%%%%%
\begin{titlepage}
\begin{flushright} {ROM2F/97/31}\\
\end{flushright}
\vskip 1mm
%%%%%%%%%%%%%%%%%%%%%%%  TITOLO  %%%%%%%%%%%%%%%%%%%%%%%%%%%%%%%%%%%%%%%%%
\begin{center}  {\large \bf Cross Section of a Resonant-Mass  Detector }\\  
{\large \bf for Scalar Gravitational Waves } \\   
\vspace{2.0cm}
%%%%%%%%%%%%%%%%%%%%%%%%  AUTORI  %%%%%%%%%%%%%%%%%%%%%%%%%%%%%%%%%%%%%%%%%  
{\bf M. Bianchi}\footnote{  {\sl  Dipartimento di Fisica, Universit\`a di Roma
II ``Tor Vergata"}}
\hspace{.1mm}\footnote{{\sl  I.N.F.N. Sezione di Roma II,  Via Della Ricerca
Scientifica, 00133 Roma, ITALY}}, {\bf M.
Brunetti}\hspace{1mm}$^{1\hspace{1mm}2}$, {\bf E.
Coccia}\hspace{1mm}$^{1\hspace{1mm}2}$, {\bf F.
Fucito}\hspace{1mm}$^{2\hspace{1mm}}$, {\bf J.A. Lobo}\footnote{{\sl Departament
de Fisica Fonamental, Universitat de Barcelona, Diagonal 647, Barcelona, Spain}}
\vskip 3.0cm
%%%%%%%%%%%%%%%%%%%%%%%%  ABSTRACT  %%%%%%%%%%%%%%%%%%%%%%%%%%%%%%%%%%%%%%  
{\large \bf Abstract}\\
\end{center} Gravitationally coupled scalar fields, originally introduced by
Jordan, Brans and Dicke to account for a non constant gravitational coupling,
are a prediction of many  non-Einsteinian theories of gravity not excluding
perturbative formulations of String Theory. In this paper, we compute the cross
sections for scattering and absorption  of scalar and tensor gravitational waves
by a resonant-mass detector in the framework of the Jordan-Brans-Dicke theory.
The results are then specialized to the case of a detector of spherical shape
and shown to reproduce those obtained in General Relativity in a certain limit.
Eventually we discuss the potential detectability of scalar waves emitted in a spherically
symmetric gravitational collapse.
\vfill
\end{titlepage}
%%%%%%%%%%%%%%%%%%%%%%%%%%%%%%%%%%%%%%%%%%%%%%%%%%%%%%%%%%%%%%%%%%%%%%%%%%
%%%%%%%%%%%%%%%%%%%%%  INIZIO TESTO %%%%%%%%%%%%%%%%%%%%%%%%%%%%%%%%%%%%
%%%%%%%%%%%%%%%%%%%%%%%%%%%%%%%%%%%%%%%%%%%%%%%%%%%%%%%%%%%%%%%%%%%%%%%%%%
\addtolength{\baselineskip}{0.3\baselineskip}
%%%%%%%%%%%%%%%%%%%%%%%%%%%%%%%%%%%%%%%%%%%%%%%%%%%%%%%%%%%%%%%%%%%%%%%%%%
%%%%%%%%%%%%%%%%%%%%%%%%%%%%%%%%%%%%%%%%%%%%%%%%%%%%%%%%%%%%%%%%%%%%%%%%%%
%%%%%%%%%%%%%%%%%%%%%%%%%%%%%%%%%%%%%%%%%%%%%%%%%%%%%%%%%%%%%%%%%%%%%%%%%%

%%%%%%%%%%%%%%%%%%%%%%%%%%%%%%%%%%%%%%%%%%%%%%%%%%%%%%%%%%%%%%%%%%%%%%%%%%
%%%%%%%%%%%%%%%%%%%%%%  INTRODUCTION  %%%%%%%%%%%%%%%%%%%%%%%%%%%%%%%%%%%%
%%%%%%%%%%%%%%%%%%%%%%%%%%%%%%%%%%%%%%%%%%%%%%%%%%%%%%%%%%%%%%%%%%%%%%%%%% 

\section{Introduction} Possible extensions of Einstein's theory of gravitation
to include  scalar fields have received much attention in the past years. The
existence of a scalar component in the gravitational field has been originally
proposed by Jordan and independently by Brans and Dicke \cite{bd} in order to
devise a theoretical framework allowing for variations of the fundamental
constants and violations of the (Strong) Equivalence Principle. Many other
non-Einsteinian theories of gravity incorporate scalar fields. Most notably
String Theory, the most serious candidate for a fully consistent quantum
theory of gravity,  generically predicts the existence of neutral scalar fields
\cite{gsw}. In particular all perturbative string vacua include a scalar, known
as the  dilaton, in their massless spectrum. Its vacuum expectation value,
which  plays the role of the string coupling constant, is neither fixed at the
classical level nor at any order in perturbation theory. Very poorly understood
non-perturbative effects may generate a potential for the dilaton and the other
scalar fields thus stabilizing their expectation value. This mechanism is by no
way incompatible with some scalars remaining massless \cite{dp}. The active
research in this field gives new motivations for further investigating theories
of gravitation including other scalars (dilaton, axions or the superpartners of
the known fermions). In this respect, it is worth observing that, assuming the
validity of a least coupling principle \cite{dp}, the subsector of String
Theory determining the coupling of the dilaton has the same functional form
of the Jordan-Brans-Dicke (JBD) theory. 

Most probably, the existence of massless gravitationally coupled scalar fields
would   be detected through deviations from General Relativity (GR) in the
spin contents of gravitational waves (GWs). In this respect, one of the most
promising sources of GWs is given by the gravitational collapse of a star
\cite{kt}. Since in GR no gravitational radiation is emitted in a spherically 
symmetric collapse, most of the existing literature focuses on the
non-spherically symmetric case which generates GW of spin two. However, in
scalar-tensor theories, scalar  gravitational waves are radiated from a
spherically symmetric collapse. In the Oppenheimer-Snyder approximation, such
emission process has been recently studied \cite{harada}. Theoretical
predictions for the amplitude of GWs depend on the specific model chosen to
describe the collapse and on the assumed theory of gravity. The only constraint
is that the assumed theory of gravity must agree, in the regime of weak
gravitational fields, with the existing experimental data \cite{will} which
support GR to a high degree of accuracy. In the regime of strong gravitational
fields the situation is different and large deviations from GR are possible in
principle \cite{de}. Eventually, we will argue that scalar GWs emitted in a
spherically  symmetric collapse in the strong field regime could give a
measurable effect for a source within our local group of galaxies.

Among the GW detectors which are now under study or in construction, those
with spherical symmetry \cite{cf, clo, jm, mz} are in a priviledged position
\cite{ad, wp} to detect and discriminate scalar waves. 
Neither a single cylindrical-shape
resonant-mass detector nor a single laser interferometer are in fact able to
perform this task. A proposed omnidirectional observatory made out of six
cylindrical   resonant-mass detectors \cite{cal} should be able to 
discriminate the
scalar component of a GW. A minimum of four laser
interferometers are needed to  discriminate the scalar mode \cite{mshi}. On the
contrary a single  spherical resonant-mass  detector was shown to be able
to detect and discriminate a scalar mode \cite{wp,lobo} and to act as a veto for
different theories of gravity
\cite{bccff}. This can be accomplished by monitoring the five degenerate
fundamental  quadrupole modes of vibration of the detector together with the
fundamental monopole mode. In fact, in any metric  theory of gravity the
``electric" component of the Riemann tensor $R_{0i0j}$ can be written (in the
so-called Jordan-Fierz frame) as
\be  R_{0i0j}=S_{ij}+\frac{1}{3}\,T\delta_{ij} \quad ,
\label{uno}
\ee
where $S_{ij}$ is a traceless symmetric tensor, and $T=R_{0i0i}$ is the 
trace part\footnote{
A convenient basis for symmetric rank two tensors is 
\{$S^{ij}_{(2m)}$; $S_{(00)}^{ij}$\} \protect\cite{lobo, bccff}, which
allows to express the spherical harmonics with $l=0$ and $l=2$ as 
$Y_{(lm)}=S^{ij}_{(lm)}\hat x_i \hat x_j$.}. From measurements of the above
(quadrupole and monopole) modes one is able to completely reconstruct
$R_{0i0j}$ \cite{lobo, bccff}.  

In order to make quantitative predictions about the  possibility of detecting
scalar GWs, we compute in this paper the cross section for scattering and
absorption of scalar and tensor GWs by a spherical resonant-mass detector
in the framework of the JBD theory. We then apply the results to estimate the
potential ability of such a GW detector to sense the characteristic signal
emitted in the process of a spherically symmetric stellar collapse.

\section{Scalar and Tensor GWs in the JBD Theory}

Scalar particles may be coupled to gravity in many ways consistent with General
Covariance. The experimental tests of the Equivalence Principle put however
severe constraints and tend to favour theories that predict a universal 
coupling of such scalar particles to the rest of matter fields \cite{dp}.
For a  single  scalar, that may be thought as the string dilaton, the
relevant couplings may be encoded in the JBD theory. In the Jordan-Fierz
frame, in which the scalar mixes with the metric but decouples from matter,
the action reads 
\bea S &=& S_{\rm grav}[\phi,g_{\mu\nu}]+S_{\rm m}[\psi_m,g_{\mu\nu}]
= \nonumber\\ &=& {c^3\over 16\pi}\int d^4x\sqrt{-g}\left[\phi R-
{\omega_{_{BD}}\over\phi}g^{\mu\nu}\partial_\mu\phi\partial_\nu\phi\right]
+{1\over c} \int d^4x L_{\rm m}[\psi_{\rm m},g_{\mu\nu}] \quad ,
\label{due}
\eea
where $\omega_{_{BD}}$ is a dimensionless constant, whose lower bound
is fixed to be $\omega_{_{BD}}\approx 600$ by experimental data \cite{mssr},
$g_{\mu\nu}$ is the metric, $\phi$ is a scalar field, and $\psi_{\rm m}$
collectively denotes the matter fields of the theory. The part of the
Lagrangian which describes the matter sector does not depend on the field
$\phi$, and it is the same as in GR. Notice that a Weyl rescaling of the
metric, $g_{\mu\nu}=\phi^{-1/2}g^E_{\mu\nu}$, brings the gravitational part
of the action to the standard Einstein-Hilbert form but introduces direct
couplings of the scalar field $\f$ to matter. In order to perform our
computations the Jordan-Fierz frame proves to be more convenient than
Einstein's. The independence of the physical results on the frame choice
can be explicitly checked.

As a preliminary analysis, we perform a weak field approximation around the
Minkowskian metric and a constant expectation value for the scalar field 
\bea  g_{\mu\nu}&=&\eta_{\mu\nu}+h_{\mu\nu}\nonumber \\
\f &=&\f_0+\xi \quad .
\label{tre}
\eea
The standard parametrization $\f_0=2(\o_{_{BD}}+2)/G(2\o_{_{BD}}+3)$, with
$G$ the Newton constant, reproduces GR in the limit
$\omega_{_{BD}}\rightarrow\infty$ which implies $\f_0\rightarrow {1/G}$. The
linearized field equations which correspond to the variation of  (\ref{due})
with respect to $g_{\mu\nu}$ are then given by
\bea  & &-{1\over 2}\,(\p_\a\p^\a h_{\mu\nu}-
\p^\a\p_{(\mu} h_{\nu)\a}+\p_\mu\p_\nu h)+{1\over 2}\,\eta_{\mu\nu}(\p_\a\p^\a
h -\p^\a\p^\b h_{\b\a})=\nn \\ &=&{8\pi\over c^4\f_0}\,T_{\mu\nu} +
{1\over\f_0}\,[\p_\mu\p_\nu\xi-\eta_{\mu\nu}\p_\a\p^\a\xi] 
\label{quattro}
\eea
where $h$ is the trace of the fluctuation $h_{\mu\nu}$ and $T_{\mu\nu}$ is
the matter stress-energy tensor. Defining the new field
\be
\th_{\mu\nu}= h_{\mu\nu}-{1\over 2}\eta_{\mu\nu}h-\eta_{\mu\nu}{\xi\over \f_0}
\quad ,
\label{sei}
\ee and choosing the gauge
\be  
\p_\mu \th^{\mu\nu}=0 \quad ,
\label{cinque}
\ee yield the final form of the linearized field equations
\bea \label{sette} 
\p_\a\p^\a\th_{\mu\nu}&=&-{16\pi\over \f_0}T_{\mu\nu}\quad,\\
\p_\a\p^\a\xi &=&{8\pi\over 2\o_{_{BD}}+3}T \quad .
\label{settebis} 
\eea Far from the sources these equations admit wave--like solutions
\bea
\th_{\mu\nu}(x)&=& A_{\mu\nu}(\vec x, \o)\exp(ik^\a x_\a)+ c.c. \label{otto} \\
\xi(x)&=&B(\vec x, \o)\exp(ik^\a x_\a)+ c.c. \label{ottobis}
\eea Without affecting the gauge condition (\ref{cinque}), we can impose
$h=-2\xi/\f_0$ (so that
$\th_{\mu\nu}=h_{\mu\nu}$). Gauging away the superflous components, we can
write the amplitude
$A_{\mu\nu}$ in terms of the three degrees of freedom associated to states with
helicities $\pm 2, 0$ \cite{lee}. For a wave travelling in the $z$-direction,
we thus obtain
\be   A_{\mu\nu}=\pmatrix{0&0&0&0\cr 0&e_{11}-b&e_{12} &0\cr 0&e_{12}
&-e_{11}-b&0\cr 0&0&0&0\cr},
\label{nove}
\ee where $b=B/\f_0$.

\section{Cross Sections for Resonant Mass Detectors}

Before performing the computation of the cross section we would like to clearly
state the nature of our approximations. We consider GW emitted from a distant
source. For the purposes of our computations, we are not interested in the
details of the emission but we assume that the GW has the form given in
(\ref{otto}), (\ref{ottobis}) with a frequency $\omega= c |{\vec k}|$
coincident with one of the vibrational eigenfrequencies of the detector. When
the GW impinges on the resonant-mass detector, a part of the GW gets scattered
and the rest is absorbed. The size of the detector, $R$, is such that
$R\ll\lambda$, where $\lambda$ is the GW wavelenght, so that the interaction
is point--like (``quadrupole approximation''). Once excited, the detector
re--emits part of the absorbed radiation, while the rest is transformed into
noise. 

In the following we compute the scattering cross section
\be
\s_{scat} := {P_{scat}\over \Phi}\qquad,
\label{scat}
\ee
where $\Phi$ is the incident GW energy per unit time and unit area, and 
$P_{scat}$ is the power subtracted by the scattered wave, and the total
cross section
\be
\s_{tot}:={P_{scat} + P_{abs}\over \Phi} = -{P_{int}\over \Phi}\qquad,
\label{tot}
\ee
where $P_{abs}$ is the power absorbed by the detector and $P_{int}$ is the
power associated to the interference between the incident and the scattered
wave \cite{wei}.

All the computations will be performed twice: once for the tensorial waves
(\ref{otto}) and once for the scalar waves (\ref{ottobis}). We note that the
GW given by (\ref{otto}) receives contributions both from a traceless tensor
term and from a scalar term, the trace of $\theta_{\mu\nu}$. The computation
of the traceless part is similar to the one performed in the context of
standard GR, which is recovered in the limit
$\omega_{_{BD}}\rightarrow\infty$. Since the computation in the latter case
si very well-known, we will parallel it, describing the general framework but
omitting some details which can be found in \cite{wei}. All the formalism
developed will then be applied to the scalar case.

At large distances from the detector, $r=|\vec x| \to\infty $, the GW is a
superposition of a plane wave and a scattered wave
\be 
\th_{\mu\nu}(\vec x, t)\to 
\left[A_{\mu\nu}e^{ik\cdot x}+ H_{\mu\nu}(\hat x){e^{i\o r}\over
r}\right]e^{-i\o t}+ c. c.\quad,
\label{dieci}
\ee
where $H_{\mu\nu}$ is the scattering amplitude. Expanding the plane wave in
spherical waves, we get
\be e^{ik\cdot x}\to {e^{i\o r}\over i\o r}\d(1-\hat k\cdot\hat x)-
{e^{-i\o r}\over i\o r}\d(1+\hat k\cdot\hat x)\quad,
\label{undici}
\ee
where $\hat k$ and $\hat x$ are the unit vectors in the direction of
$\vec k$ and $\vec x$ respectively. Plugging (\ref{undici}) back into
(\ref{dieci}) yields
\be
\th_{\mu\nu}\to [A_{\mu\nu}^{\rm out}\ e^{i\o r}+ A_{\mu\nu}^{\rm in}\ 
e^{-i\o r}]e^{-i\o t}+ c.c. \quad ,
\label{dodici}
\ee
where
\bea  A_{\mu\nu}^{\rm out}(\vec x) &=& {1\over i\o r}\ [A_{\mu\nu}\d(1-\hat
k\cdot\hat x)+i\o H_{\mu\nu}(\hat x)],
\label{tredici}\\ A_{\mu\nu}^{\rm in}(\vec x) &=& -{1\over i\o r}\
A_{\mu\nu}\d(1+\hat k\cdot\hat x).
\label{tredicibis}
\eea

If we choose a GW travelling in the $z$ direction, with wave vector
$k^\mu=(\o,0,0,\o)$, the perturbation $\th_{\mu\nu}$ will have non-vanishing
components only for $i,j\neq 0, z$ (see (\ref{nove})). Then, keeping into
account this choice, we can introduce the stress--energy pseudo--tensor
$t_{\mu\nu}$, that results from an expansion of the equations of motion to
second order in the weak fields. In particular the mixed components read
\be   <t_{0z}> = -\hat z{\f_0 c^4\over 32\pi}
\left[{4(\o_{_{BD}}+1)\over \f_0^2}<(\p_0\xi)(\p_0\xi)>+ <(\p_0 h_{\a\b})(\p_0
h^{\a\b})>\right],
\label{quattordici}
\ee
where the symbol $<...>$ implies an average over a region of size much
larger than the wavelength of the GW.  Substituting (\ref{otto}), (\ref{ottobis})
into (\ref{quattordici}) we get
\be   <t_{0z}> = -\hat z {\f_0 c^4 \o^2 \over 16\pi}
\left[{2(2\o_{_{BD}}+3)\over \f_0^2} \mid B\mid^2 + A^{\a\b *}A_{\a\b}-{1\over
2}\mid {A^\a}_\a
\mid^2\right],
\ee
and using (\ref{nove}), we obtain
\be  <t_{0z}>=-\hat z {\f_0 c^4 \o^2\over 8\pi}
\left[\mid e_{11}\mid^2+\mid e_{12}\mid^2+ ({2\o_{_{BD}}+3})\mid b \mid^2\right].
\label{quattordicibis}
\ee From (\ref{quattordicibis}) we see that the purely scalar contribution,
associated to $b$, and the traceless tensorial contribution, associated to
$e_{\mu\nu}$, are completely decoupled and can thus be treated independently.

\subsection{Cross Section for Tensor GWs}

For spin--two waves, the scattering cross section is given by \cite{wei} 
\be 
\s_{scat} = {\int \left[f^{\l\nu *}f_{\l\nu}-{1\over 2}\mid {f_\a}^\a
\mid^2\right] d\O
\over \left[e^{\l\nu *}e_{\l\nu}-{1\over 2}\mid {e_\a}^\a
\mid^2\right]}, 
\label{ventidue}
\ee
where $f_{\mu\nu}$ is the spin--two component of the scattering amplitude
$H_{\mu\nu}$. The total cross section is given by \cite{wei}
\be 
\s_{tot} = {4\pi\Im\left\{e^{\l\nu *}f_{\l\nu}(\hat k)- {1\over 2}{e^{\b*}}_\b
{f^\a}_\a(\hat k)
\right\}\over
\o \left[e^{\l\nu *}e_{\l\nu}-{1\over 2}\mid {e^\a}_\a \mid^2\right]}. 
\label{ventitre}
\ee
The ratio between the two
\be
\eta = {\s_{scat} \over \s_{tot}}
\label{ventinove}
\ee exactly coincides with the ratio between the energy re--emitted from the
resonant detector as GWs and the energy transformed into noise, viz.,
\be
\eta={1\over \G}{P_{GW}\over E_{osc}},
\label{ventotto}
\ee
where $1/\Gamma$ is the decay time of the free oscillation of the detector.

Which particular combination of vibrational modes of the detector gets excited
clearly depends on the polarization of the incoming GW. In the
``qua\-dru\-po\-le approximation'' the scattering  amplitude can be
expressed as
\be f_{\mu\nu}(\hat x)=
\tau_{\mu\nu}(\hat x)-{1\over 2}\eta_{\mu\nu}\tau(\hat x).
\label{ventiquattro}
\ee
where $\tau_{\mu\nu}(\hat x)$ are (proportional to) the Fourier transform of
the stress-energy tensor at $\vec k^\prime = \omega \hat x$. Stress-energy
conservation allows one to recast (\ref{ventidue}) into 
\be 
\s_{tot}={2\pi\Im
\{e^*_{11}(\tau_{11}-\tau_{22})+ 2e^*_{12}\tau_{12}\}\over
\o [\mid e_{11} \mid^2+\mid e_{12}\mid^2]} \quad ,
\label{ventisei}
\ee that depends only on the traceless components of $e_{\mu\nu}$ and
$\tau_{\mu\nu}$, and (\ref{ventitre}) into
\be 
\s_{scat}={4\pi\left[\tau^*_{ik}\tau_{ik}- {1\over 3}\mid
\tau_{ii}\mid^2\right]\over 5 [\mid e_{11}
\mid^2+\mid e_{12} \mid^2]}.
\label{ventisette}
\ee

For a ``pointlike" spherical detector $(R\ll\lambda)$
\be 
\tau_{jk}(\o)=\g e_{jk}
\label{trenta}
\ee where $\g$ is a constant to be determined shortly. By introducing a set of
five matrices
$S_{(2m)}^{jk}$ which form a convenient basis for the spherical harmonics with
$l=2$ \cite{lobo}, and choosing $+z$ as the direction of propagation, we obtain
\be \label{modiGR}
\tau^{jk}(\o)=\g \sqrt{8\pi\over 15} [e_{11}(S_{(22)}^{jk}+S_{(2-2)}^{jk})-i
e_{12}(S_{(22)}^{jk}-S_{(2-2)}^{jk})]
\ee From (\ref{modiGR}), it is clear that a tensorial GW propagating in the
$+z$-direction will excite only two modes of the detector, precisely those
with $l=2$ and $m=\pm 2$. Plugging (\ref{trenta}) into (\ref{ventisei}) and
(\ref{ventisette}), the condition (\ref{ventinove}) becomes 
\be 
\Im (\g)={2\o\over 5\eta}\mid\g\mid^2,
\label{trentuno}
\ee
which allows us to determine $\g$ in terms of $\eta$ and $\o$. Moreover, at
resonance ($\o\simeq\o_2$) the modes of $T^{\mu\nu}$  behave as damped
harmonic oscillators. Fourier transforming, one easily infers the
$\o$-dependence of $\tau^{\mu\nu}$, and finally gets 
\be 
\s_{tot}=\left({10\pi\eta c^2\over \o^2}\right) {\G^2/4\over
(\o-\o_2)^2+\G^2/4}, 
\label{trentadue}
\ee
where $\o_2$ is the resonance frequency of one of the quadrupole mode
($l=2$) of the detector. The eigenfrequencies $\o_{nl}$ can be simply labeled
by the radial quantum number $n$ and the principal angular quantum number $l$,
since by spherical symmetry they do not depend on the azimuthal quantum number
$m$, i.e., they are (2$l\/$+1)-fold degenerate. A few numerical values of the
eigenfrequencies $\o_{nl}$ of the spheroidal modes can be found, for example,
in \cite{lobo,bccff}\footnote{
Toroidal modes of a spherical detector cannot be excited by GW in any metric
theory, and can thus be used as a veto in the detection \protect\cite{bccff}.}.
The last task we have to perform is the computation of $\eta$ from
(\ref{ventotto}). To this end we need the power emitted as GWs from the
detector. For the spin--two components under consideration,
\be  P_{GW}={2\o^6\over 5\f_0 c^5} D^{ij*}_{(T)}(\o)D_{ij(T)}(\o). 
\label{trentaquattro}
\ee
The traceless quadrupole moment is defined as
\be   D^{ij}_{(T)}(\o) = D^{ij}(\o)-{1\over 3}\d^{ij} D_k{}^k(\o), 
\label{trentacinque}
\ee
where  
\be D^{ij}(\o) := \int x^i x^j \rho(\vec x, \o)\ d\vec x 
\label{trentacinquebis}
\ee
is the quadrupole moment of the detector and $\rho$ is the mass density. 
This quadrupole moment is due to
the mass variation of the detector forced by the incoming GW. Its computation
is not particularly enlightening, and we report here only the final result for
a spherical detector of radius $R$:
\bea 
& & D^{ij}_{(T)}(t)= {16\pi\over 15}\rho R^4 C(n, 2) e^{-i\o_{n2} t} 
\sum_m S^{ij}_{(2m)}\times \nn\\ &\times &\left[\b_3(k_{n2}R)
{j_2(q_{n2}R)\over q_{n2}R}-3{q_{n2}\over k_{n2}}\b_1(q_{n2}R)
{j_2(k_{n2}R)\over k_{n2}R} \right]+ c.c.=\nn \\ &=& D^{ij}(t) +
{8\pi\over 3} \rho R^4 C(n, 0) e^{-i\o_{n0}t} S^{ij}_{(00)}
\left[\b_3(k_{n0}R) {j_2(q_{n0}R)\over q_{n0}R}\right]+c.c.  
\label{trentasei}
\eea
where $j_l(x)$ are spherical Bessel functions \cite{as}, and 
\be q^2_{nl}={\rho\o_{nl}^2\over\l+2\mu};\qquad  k_{nl}={\rho\o_{nl}^2\over\mu}
\label{trentasette}
\ee
enforce the dependence on the material used to build the detector through
the Lam\'e coefficients $\l, \mu$.  The auxiliary functions $\b_i$'s are 
\bea
\b_1(z) & := &{d\over dz}\left({j_l(z)\over z}\right)\nn \\
\b_2(z) & := &{d^2 j_l(z)\over dz^2}\nn \\
\b_3(z) & := &{1\over 2}\left[\b_2(z)+ (l+2)(l-1){j_l(z)\over z^2}\right].
\label{trentotto}
\eea
The normalization constants $C(n,l)$ are given by \cite{lobo,brunm}
\be
\mid C(n, l)\mid^2 = {4\pi\over 3} (k_{nl}R)^3
\left\{\int_0^{k_{nl} R} [F_{1(nl)}(r)^2 + l(l+1) F_{2(nl)}(r)^2] d(k_{nl}
r)\right\}^{-1}
\label{trentanove}
\ee where
\bea  F_{1(nl)}(r) &=&
\b_3(k_{nl} R)k_{nl} r {d\over d(q_{nl} r)}j_l(q_{nl} r)- l(l+1){q_{nl} \over
k_{nl}}\b_1(q_{nl}R) j_l(k_{nl} r)\nn \\ F_{2(nl)}(r)&=&{k_{nl}\over
q_{nl}}\b_3(k_{nl}R)j_l(q_{nl} r) -{q_{nl}\over k_{nl}}\b_1(q_{nl} R) {d\over
d(k_{nl} r)} [k_{nl} r j_l (k_{nl} r)].
\label{quarantabis}
\eea
A more detailed exposition of the above computations can be found in
\cite{brunm}.

Substituting (\ref{trentasei}) into (\ref{trentaquattro}) yields
\bea P_{GW}&=&
% {64\pi\over 15 \f_0 c^5} \rho^2 R^8 \mid C(n, 2)\mid^2\
% \o_{n2}^6\nn \\ & &\left[\b_3(k_{n2}R){j_2(q_{n2}R)\over q_{n2}R}-
% 3{q_{n2}\over k_{n2}}\b_1(q_{n2}R){j_2(k_{n2}R)\over k_{n2}R}\right]^2
% =\nn \\ &=&
{12\over 5\pi \f_0 c^5} M^2 R^2\,|C(n, 2)|^2\,\o_{n2}^6\ \times \nn \\
& \times &\left[\b_3(k_{n2}R){j_2(q_{n2}R)\over q_{n2}R}-
3{q_{n2}\over k_{n2}}\b_1(q_{n2}R){j_2(k_{n2}R)\over k_{n2}R}\right]^2, 
\label{quaranta}
\eea
where $M\/$ is the total mass of the sphere. The oscillation energy of the
five modes with $l=2$ is given by
\bea  E_{osc}^{(n)} &=& {15\over 2\pi}\,M\o_{n2}^2\,|C(n, 2)|^2
{1\over (k_{n2} R)^3}\ \times \nn \\
& \times &\int_0^{k_{n2} R}\{F_{1(n2)}(r)^2+6 F_{2(n2)}(r)^2\}\,d(k_{n2} r).
\label{quarantuno}
\eea
Making use of (\ref{quaranta}) and (\ref{quarantuno}), we can find the
explicit value of (\ref{ventotto}) which, inserted into (\ref{trentadue}),
leads to the final expression for the total cross section of a spin--two GW
by a spherical detector in the context of JBD theory:
\be  
\s_{tot}^{(n)}= F_n\,{G M v_s^2\over c^3}\,{2\o_{_{BD}}+3\over 2(\o_{_{BD}}+2)}
\,{\G \over (\o-\o_{n2})^2+\G^2/4},
\label{quarantadue}
\ee where
\be  v_s=\sqrt{2(1+\s_{_P})}\,{\o_{nl}\over k_{nl}},
\ee   is the sound velocity, 
\be 
\s_{_P}= {\l\over 2(\mu+\l)}
\ee is the Poisson ratio, and finally 
\bea  F_n & := &{2\pi\over 5(1+\s_{_P})}
\left[\b_3(k_{n2}R){j_2(q_{n2}R)\over q_{n2}R}- 3{q_{n2}\over
k_{n2}}\b_1(q_{n2}R) {j_2(k_{n2}R)\over k_{n2}R}\right]^2\ \times \nn \\
& \times &(k_{n2} R)^5\left\{\int_0^{k_{n2} R}[F_{1(n2)}(r)^2+
6 F_{2(n2)}(r)^2]\,d(k_{n2}r)\right\}^{-1}.
\label{quarantatre}
\eea It is useful to compute the integrated cross section
\be
\Sigma_n={1\over 2\pi}\int^{+\infty}_{-\infty}\sigma_{tot}^{(n)} d\o =
{GMv_s^2\over c^3}{ F_n (2\o_{_{BD}} + 3) \over 2(\o_{_{BD}} + 2} .
\label{tensintcross}
\ee A few numerical values of $F_n$ are given in Table 1 for a standard
value of the Poisson ratio, $\s_{_P}=1/3$. For this value of $\s_P$, an
analytic expression for the integral which appears in the definition of
$F_n$ is given in the appendix. Note that (\ref{quarantadue}) correctly
reproduces the GR result \cite{wp,tdc} in the limit
$\omega_{_{BD}}\rightarrow\infty$.

\subsection{Cross Section for Scalar GWs}

We now turn to a detailed computation of the cross section for a scalar GW.
We begin by determining the energy flux of the incoming scalar waves:
~(\ref{quattordicibis})
\be
\Phi_{(s)}=\hat x_k <t^{0k}>_{(s)}={\o^2 c^4\over 8\pi} {2\o_{_{BD}}+3\over
\f_0}\mid B\mid^2,
\label{quarantaquattro}
\ee where the subscript $s$ stands for ``scalar''. If we denote the scattering
amplitude by $W$, at large distances from the detector, and in complete analogy
with (\ref{dieci}) the scalar GW is a superposition of a plane and a scattered
wave:
\be
\xi(\vec x, t)\to \left[B e^{ik\cdot x}+ W(\hat x){e^{i\o r}\over
r}\right]e^{-i\o t} + c.c.
\label{quarantacinque}
\ee
We can thus define incoming and outgoing amplitudes:
\bea B^{\rm out}(\vec x) &=& {1\over i\o r}\ [B \d(1-\hat k\cdot\hat x)+
i\o W(\hat x)], \label{quarantasei}\\
B^{\rm in}(\vec x) &=& -{1\over i\o r}\ B \d(1+\hat k\cdot\hat x),
\label{quarantaseibis}
\eea
By substituting (\ref{quarantasei}) into (\ref{quarantaquattro}) one can
compute the power associated to the outgoing part of the GW. The interference
between the incident plane wave and the scattered wave leads to a contribution
\be P_{int} = -{\o\over 4\pi\f_0}(2\o_{_{BD}}+3)
\Im\left\{\int d\O\ \d(1-\hat k\cdot\hat x) B^* W(\hat x)\right\},
\label{quarantasette}
\ee while the contribution of the scattered wave is
\be P_{scat} = {\o^2\over 8\pi\f_0}(2\o_{_{BD}}+3)\int d\O\ |W|^2.
\label{quarantotto}
\ee The scattering cross section is 
\be
\s_{scat} := \left({P_{scat}\over \Phi}\right)_{s}=
{\int d\O\ |W|^2\over |B|^2},
\label{quarantanove}
\ee
and the total cross section is
\be
\s_{tot}:=-\left({P_{int}\over\Phi}\right)_{s}=
{2\over \o}\,{\Im\left\{\int d\O\
\d(1-\hat k\cdot\hat x) B^* W(\hat x)\right\}\over |B|^2}.
\label{cinquanta}
\ee
The ``quadrupole approximation'' and the conservation of the stress--energy
tensor  imply $T_{ij}(\vec k, \o) = - {\o^2 \over 2} D_{ij}(\o)$ and allow
one to express the incoming wave amplitude as
\bea  B(\vec x, \o)&\simeq &-{2\over (2\o_{_{BD}}+3)r c^4} \int T(\vec x',\o)
e^{-i\vec k\cdot\vec x'}\ d\vec x'\ =\nn\\ &=&-{2\over (2\o_{_{BD}}+3)r c^4}
T(\vec k,\o)\ =\nn\\ &=&-{2\over (2\o_{_{BD}}+3)r  c^4}[T_{jj}(\vec k,\o)-
T_{00}(\vec k,\o)]\ =\nn\\ &=&-{2\over (2\o_{_{BD}}+3)rc^4}[T_{jj}(\vec k,\o)
-\hat x_j\hat x_k T_{jk}(\vec k, \o)]\ =\nn\\ &\simeq & {\o^2\over 
(2\o_{_{BD}}+3)r c^4} [\d_{jk}-\hat x_j\hat x_k] D_{jk}(\o).
\label{cinquantuno}
\eea In analogy with the form of (\ref{cinquantuno}), the scattering amplitude
can be written as
\be W(\hat x)={\tau_\b}^\b(\hat x)=(\d_{jk}-\hat x_j \hat x_k)\tau_{jk}
\label{cinquantadue}
\ee and, once substituted into (\ref{quarantanove}), (\ref{cinquanta}), yields
\bea  \label{cinquantatre}
\s_{scat}&=&{8\pi\over 5} {|\tau_{ii}|^2+\tau^{ij*}\tau_{ij}/3\over |B|^2},\\
\s_{tot}&=&
% {2\over \o} {\Im\left\{\int d\O\ \d(1-\hat k\cdot\hat x)
% B^*(\tau_{ii}-\hat x_i\hat x_j\tau_{ij})\right\} \over |B|^2}=\nn\\ &=&
{4\pi\over \o}\,{\Im\{B^*  (\tau_{ii}-\hat k_i\hat k_j \tau_{ij})\}
\over |B|^2}.
\label{cinquantatrebis}
\eea

Like in the spin--two case, the vibrational modes of the detector which are
excited by an incoming GW depend on the polarization of the  GW. Thus, in the
case of a scalar GW propagating in the $+z$ direction, the excited modes are
those with \{$l=m=0$\} {\it and also\/} those with \{$l=2$, $m=0$\}. This is
because the space components of the trace part of the GW tensor (\ref{nove})
must be expressed as a linear combination of $S_{(00)}^{jk}$ and
$S_{(20)}^{jk}$ \cite{lobo}. For a spherically shaped detector, the
eigenfrequencies corresponding to the spheroidal modes with quantum numbers
$l=m=0$ ($\o_0$) and $l=2,m=0$ ($\o_2$) are numerically different \cite{wp,
lobo, bccff}, and consequently we have to consider two cases: the scattering
amplitudes for a GW travelling in the $z$-direction are given by 
\bea
\tau^{ij}(\o)&=&\a B S_{(00)}^{ij}
\qquad\qquad\hbox{for}\qquad\o\simeq\o_0\nn\\
\tau^{ij}(\o)&=&\b B S_{(20)}^{ij}
\qquad\qquad\hbox{for}\qquad\o\simeq\o_2,
\label{cinquantaquattro}
\eea which, once substituted into (\ref{cinquantatre}) and
(\ref{cinquantatrebis}), in conjunction with (\ref{ventinove}),
lead to the conditions 
\be
\Im(\a)={2\o\over\eta_0\sqrt{4\pi}}\mid\a\mid^2\qquad
\Im(\b)=-{\o\over 2\eta_2\sqrt{5\pi}}\mid\b\mid^2.
\label{cinquantacinque}
\ee
At resonance ($\o\simeq \o_2$ or $\o\simeq \o_0$) the modes of $T^{\mu\nu}$ 
behave as damped harmonic oscillators. Fourier transforming, one easily
obtains the $\o$-dependence of $\tau^{\mu\nu}$, and finally 
\bea
\a &=& {\eta_0\sqrt{4\pi}\over 2\o}\left({-\G_0/2\over 
\o-\o_0+i\G_0/2}\right)\nn\\
\b &=& {2\eta_2\sqrt{5\pi}\over \o}
\left({\G_2/2\over \o-\o_2+i\G_2/2}\right),
\label{cinquantasei}
\eea The cross sections are thus given by
\bea  \label{cinquantasette}
\s_{tot\ (00)}&=&\left({4\pi\eta_0 c^2\over \o^2}\right) {\G_0^2/4 \over
(\o-\o_0)^2+\G_0^2/4} \\
\s_{tot\ (20)}&=&\left({20\pi\eta_2 c^2\over \o^2}\right) {\G_2^2/4 \over
(\o-\o_2)^2+\G_2^2/4}.
\label{cinquantasettebis}
\eea where $\G_0$, $\G_2$ are the decay times of the free oscillation of the
detector's modes with $l=0$ and $l=2$, and
$\eta_0$, $\eta_2$ are defined as in (\ref{ventotto}). Note the geometrical
ratio $5:1$, related to the degeneracy of the quadrupole modes, between
(\ref{cinquantasettebis}) and (\ref{cinquantasette}) for a hypothetical
detector with $\eta_0=\eta_2$, $\Gamma_0=\Gamma_2$ and $\o_0=\o_2$.

The last thing which is left to do is the computation of the  parameters
$\eta_0,\eta_2$ in (\ref{cinquantasette}), (\ref{cinquantasettebis}). Using
(\ref{trentasei}) we find the power emitted by the detector due to the
presence of the scalar field:
\bea  P_{GW} &=& {\o^2 r^2 c^3(2\o_{_{BD}}+3)\over 8\pi\f_0}\int|B(\o)|^2
d\O\ =\nn \\ &=& {\o^6\over 8\pi\f_0(2\o_{_{BD}}+3)c^5} D^{ij*}(\o)D^{lm}(\o)
\int(\d_{ij}-\hat x_i\hat x_j)(\d_{lm}-\hat x_l\hat x_m)d\O\ =\nn \\ &=&
{\o^6\over  5\f_0(2\o_{_{BD}}+3)c^5}
\left[\mid D_{jj}(\o)\mid^2+{1\over 3}D^{lm*}(\o)D_{lm}(\o)\right]\ =\nn \\
&=& {2\over 5\pi\f_0(2\o_{_{BD}}+3)c^5}M^2 R^2(5P_{00}+P_{20}),
\label{cinquantotto}
\eea where
\bea P_{00} &=& \mid C(n, 0)\mid^2\ \o_{n0}^6\
[\b_3(k_{n0}R)(j_2(q_{n0}R)/q_{n0}R)]^2 \nn \\ P_{20} &=& \mid C(n, 2)\mid^2\
\o_{n2}^6\ [\b_3(k_{n2}R)(j_2(q_{n2}R)/ q_{n2}R)\ - \nn \\ &-& 3(q_{n2}/
k_{n2})\b_1(q_{n2}R)(j_2(k_{n2}R)/ k_{n2}R)]^2.
\label{cinquantanove}
\eea

For the mode with quantum numbers $l=0,m=0$, putting together the first term
in (\ref{cinquantotto}) and the oscillation energy
\bea  E_{osc}^{(n0)}&=&{3\over 2\pi} M^2\o_{n0}^2 {\mid C(n, 0)\mid^2\over
(k_{n0}R)^3} \b_3(k_{n0}R)^2\ \times \nn \\
& \times &\int_0^{k_{n0}R}[k_{n0}rj_0'(q_{n0}r)]^2 d(k_{n0}r),
\label{sessanta}
\eea  one finds
\bea
\eta_0 &=& \left({P_{GW}\over \G_0 E_{osc}}\right)_{(00)}=\nn \\
&=&{2 G M v_t^2\o_{n0}^2\over 3\G_0(\o_{_{BD}}+2)c^5}\,
[(j_2(q_{n0}R)/q_{n0}R)]^2\ \times \nn \\
& \times & (k_{n0}R)^5 \left\{\int_0^{k_{n0}R}[k_{n0}r j'_0(q_{n0}r)]^2\,
d(k_{n0}r)\right\}^{-1},
\label{sessantuno}
\eea
where $v_t=\o_{nl}/k_{nl}$. The total cross section for resonant scattering
and absorption at $\o \approx \o_{no}$ of scalar GWs by a spherical detector
is then 
\bea
\s_{tot}^{(n0)}&=& {2\pi\over 3} {G M v_t^2\over c^3(\o_{_{BD}}+2)}
[(j_2(q_{n0}R)/q_{n0}R)]^2 (k_{n0}R)^5\ \times \nn \\
& \times & \left\{\int_0^{k_{n0}R}[k_{n0}r j'_0(q_{n0}r)]^2\,
d(k_{n0}r)\right\}^{-1}{\G_0\over (\o-\o_{n0})^2+\G_0^2/4}\ = \nn \\
& = &H_n\ {G M v_s^2\over c^3(\o_{_{BD}}+2)}\,
{\G_0\over (\o-\o_{n0})^2+\G_0^2/4},
\label{sessantadue}
\eea with
\bea H_n & := &{\pi\over 3(1+\s_{_P})}\,[(j_2(q_{n0}R)/q_{n0}R)]^2
(k_{n0}R)^5\ \times \nn \\
& \times &\left\{\int_0^{k_{n0}R} [k_{n0}r j'_0(q_{n0}r)]^2\
d(k_{n0}r)\right\}^{-1} = \nn \\
&=& {\pi\over 3(1+\s_{_P})}\,[(j_2(q_{n0}R)/q_{n0}R)]^2(k_{n0}R)^2
(q_{n0}R)^3\ \times \nn \\
& \times &\left[{1\over 2}(q_{n0}R)+{1\over 4}\sin(2q_{n0}R)-
{\sin^2(q_{n0}R)\over q_{n0}R}\right]^{-1}.
\label{sessantatre}
\eea
Taking a standard value for the Poisson ratio, $\s_{_P}=1/3$, we report
in Table~\ref{valorifnhn} the values of $H_n$ and $F_n$. It is useful to
determine also the integrated cross section:
\be 
\Si_{n0}={1\over 2\pi}\int_{-\infty}^{+\infty} \s_{tot}^{(n0)} d\o=  
{GMv_s^2\over c^3} {H_n \over \o_{_{BD}}+2}
\label{sezint_00}
\ee 

For the other mode, with quantum numbers $l=2, m=0$, using (\ref{quarantuno})
and the second term in (\ref{cinquantotto}) one finds 
\bea
\eta_2 &=&\left({P_{GW}\over \G_2 E_{osc}}\right)_{(20)}=\nn \\
&=& {2 G M v_t^2 \o_{n2}^2\over 75\G_2(\o_{_{BD}}+2)c^5}\ \times \nn\\
&\times &\left[\b_3(k_{n2}R){j_2(q_{n2}R)\over q_{n2}R}- 3{q_{n2}\over
k_{n2}}\b_1(q_{n2}R) {j_2(k_{n2}R)\over k_{n2}R}\right]^2\ \times \nn\\
&\times &(k_{n2}R)^5\left\{\int_0^{k_{n2}R}[F_{1(n2)}(r)^2+ 6F_{2(n2)}(r)^2]
\,d(k_{n2}r)\right\}^{-1}.
\label{sessantasette}
\eea
From this one gets the total cross section for resonant scattering and
absorption at $\o\approx\o_{n2}$ of scalar waves by a spherical detector 
\bea
\s_{tot}^{(n2)} & = & {2\pi\over 15}\,{G M v_t^2\over c^3(\o_{_{BD}}+2)}\,
{\G_2\over (\o-\o_{n2})^2+\G_2^2/4}\ \times \nn \\
& \times & \left[\b_3(k_{n2}R){j_2(q_{n2}R)\over q_{n2}R} -
3{q_{n2}\over k_{n2}}\b_1(q_{n2}R){j_2(k_{n2}R)\over k_{n2}R}\right]^2\ \times
\nn \\
& \times &(k_{n2}R)^5 \left\{\int_0^{k_{n2}R}[F_{1(n2)}(r)^2 +
6F_{2(n2)}(r)^2]d(k_{n2}r)\right\}^{-1} = \nn \\
& = & {F_n\over 6}{G M v_s^2\over c^3(\o_{_{BD}}+2)} {\G_2\over (\o-\o_{n2})^2+\G_2^2/4}.
\label{JBD_20}
\eea
where $F_n\/$ is given by equation (\ref{quarantatre}). The corresponding
integrated cross section is given by
\be 
\Si_{n2}={1\over 2\pi}\int_{-\infty}^{+\infty} \s_{tot}^{(n2)} d\o=  
{GMv_s^2\over c^3} {F_n\over 6 (\o_{_{BD}}+2)}
\label{sezint_20}
\ee 

\begin{table}[!htb]
\caption{Numerical values for $F_n$ and $H_n$}
\label{valorifnhn}
\begin{center}
\begin{tabular}{|c|c|c|}
\hline
$n$ & $F_n$ & $H_n$ \\ \hline 1 & 2.98 & 1.14 \\ 2 & 1.14 & 0.177 \\ 3 & 0.110
& 0.0741 \\ 4 & 0.0337 & 0.0408 \\ \hline
\end{tabular}\end{center}\end{table}

From Table~\ref{valorifnhn} we can infer the ratio between the integrated cross
section for the modes with $l=0,m=0$, and the integrated cross section for the
modes with $l=2,m=0$. For example for the vibrational mode with $n=1$, we find
$\Si_{10}/\Si_{12}=2.3$. As a last remark, we note that in the limit
$\o_{_{BD}}\to \infty$ we recover GR since $\s_{tot}^{(n2)}, \s_{tot}^{(n0)}\to
0$ and (\ref{quarantadue}) tends to the value reported in \cite{wp}.
The results of our calculations revise and extend some previous estimates 
of the cross sections obtained in
\cite{lobo}, and will find an interesting application to a binary system
of stars \cite{bcff}. In the next section we will briefly consider 
the case of a burst of
gravitational radiation emitted during the spherically symmetric collapse
of a cloud of dust.

\section{Detectability of Scalar Wave Signals}

Let us now use the calculated cross sections to evaluate the detectability by a
spherical detector of a possible scalar GW signal of astrophysical origin such as
a burst from a gravitational collapse.

We consider the spherically symmetric collapse of a homogeneous dust ball
(Op\-pen\-hei\-mer-Snyder approximation), whose scalar GW emission and wave
form have been recently studied \cite{harada}. The peak amplitude 
of the scalar GW in the JBD theory turns  out to be
\be b = {\xi\over \f_0} \simeq 10^{-23} \left({500\over \o_{BD}}\right)
\left({M_*\over M_\odot}\right) \left({10\ \hbox{Mpc}\over r}\right), 
\label{bamplitude}
\ee
where $M_*$ is the collapsing mass and $r\/$ is the distance from the source.
The characteristic frequency $f_c$, defined as the frequency at which the
energy spectrum of the waveform has its maximum value, is: 
\be f_c \simeq 3\cdot 10^3 \left({M_*\over M_\odot}\right)^{1/2}
\left({15\ \hbox{km}\over r_S}\right)^{3/2}\ \hbox{Hz},
\ee where $r_S$ is the equatorial radius of the stellar surface before the
collapse and is assumed to satisfy $r_S > 4 M_* G c^{-2}$. Using the above
figures, we can then estimate the possibility of detecting scalar GWs with a
spherical detector. To this end it is convenient to define the energy absorbed
by the detector's $n$-th mode 
\be
\Delta E_n = \int_0^\infty \Phi(\o) \s(\o) d\o \approx 2\pi \Phi(\o_n) \Sigma_n
\label{formula}
\ee where $\Phi(\o)$ is the incident GW energy flux per unit frequency. Using
the above computed integrated cross sections one gets
\be
\Delta E_n = {\pi K_n \Phi(\o_n) \over (2+\o_{_{BD}})} {G M v_s^2 \over c^3}
\label{ancora}
\ee where $K_n=2H_n$ for the mode with $l=0$ and 
$K_n=F_n/3$ for the mode with $l=2$, $m=0$. Using (\ref{quarantaquattro}) with
$b=B/\phi_o$ one finds
\be
\Delta E_n = {1 \over 4} M v_s^2 |b(\o_n)|^2 \o^2_n K_n
\label{pippo}
\ee The detector's signal-to-noise ratio can be defined as
\be SNR = {\Delta E_n \over \Delta E_{min}}
\label{sigtono}
\ee where $\Delta E_{min}$ is the minimum detectable energy innovation,
depending  on the detector's thermal and electronic noises. 
The theoretical bound on $\Delta E_{min}$ using linear readout systems is
fixed  by quantum mechanics to be $\hbar \o_n$.
For $SNR=1$ one gets
the  minimum detectable value of the Fourier transform of the scalar GW amplitude
\be |b(\o_n)|_{min} = 
\left( {4 \Delta E_{min} \over M v_s^2 \o_n^2 K_n} \right)^{1/2}
\label{bmin}
\ee
As usual in the case of short bursts, {\it i.e.}, bursts lasting for a time 
$\tau\approx 1/f_c$ much shorter than the detector's characteristic damping 
time, the peak amplitude $b\/$ and the Fourier 
transform  $b(\o)$ at $\o_n = 2\pi f_c$ can be related by 
$b \approx |b(\o_n)| f_c$.
The minimum detectable peak amplitude of the scalar GW is then
\be |b|_{min} \approx 
\left( {\Delta E_{min} \over \pi^2 M v_s^2 K_n} \right)^{1/2}
\label{bpeakmin}
\ee 

For instance let us consider a homogeneous spherical mass of a material with a
high sound velocity such as molibdenum, recently added to the traditional list of
materials used in GW research \cite{who}. 
In order to have $\o_{00}\approx 3$ kHz, taken as a typical value
in \cite{harada}, with $v_s$=5\,600 m/s, one has to take a detector diamater
of $1.8$ metres, hence $M\/$\,=\,31 tons. Substituting into (\ref{bpeakmin})
we see that $b_{min}=3$$\times$10$^{-22}$. From (\ref{bamplitude}), and taking
$\o_{_{BD}}=600$, we can estimate the maximum distance at which a solar mass
collapse can be observed through the emission of scalar GWs to be
$r_{max} \approx 0.3$ Mpc. This range includes several galaxies in our Local
Group. Assuming a rate of gravitational collapses of one event per 10 years
per galaxy, one may expect a resulting event rate approaching one event per
year in the detector.

\section{Summary and Conclusions}

Although Einstein's General Theory of Relativity is strongly supported by
all experimental evidence available to date, certain alternative theories
of the gravitational interaction naturally emerge out of more general
theoretical schemes, notably String Theory. It seems clear that any deviations
from the predictions of General Relativity must originate under very strong
gravity conditions, such as stellar collapses. We naturally
expect such phenomena to produce GWs, which will convey to the observer
information both on the physics of the source and on the limits of a given
theoretical model to understand that physics. One of the best known and well
developed alternative theories to GR is Jordan-Brans-Dicke's scalar-tensor
theory.

In this paper we have performed an in depth analysis of how JBD gravitational
waves interact with a spherical detector, which is particularly well suited to
reveal or set thresholds on non-GR GW amplitudes, e.g., monopole amplitudes.
This is a very specific feature of {\it spherical\/} detectors, for no other
individual GW antenna constructed or conceived so far, can possibly discriminate
quadrupole from monopole GW radiation: an array of such detectors is required
for this purpose, and this very significantly complicates detection techniques
and algorithms.

We have expressed our results under the form of GW {\it absorption cross
sections\/} for the different resonant modes of the antenna which get excited
by those waves, and succeeded in finding closed analytic formulas for them.
In particular, JBD waves excite the usual $m\/$\,=\,$\pm$2 quadrupole modes
of the spherical antenna, but they also excite the monopole mode {\it and\/}
the $m\/$\,=\,0 quadrupole mode. Since the frequencies of these modes are
different, we define suitable cross sections for the excitation of each of
them.

Cross sections are of course very useful to define the {\it sensitivity\/}
of a detector with a given level of noise, i.e., they enable an estimate of
signal to noise ratio. As a practical application, we have considered the
signal emitted during the spherically symmetric collapse of a cloud of dusty
matter ---an event which would never occur should GR be the correct theory of
gravity---, and assessed the possibilities of seeing it with the projected
future spherical detectors. With the present bounds on the JBD parameter
$\o_{BD}$, we conclude that such events as this could be observed if they
happen within our Local Group of galaxies, with an event rate of a rather
encouraging one per year.

The possibility of sensing or thresholding monopole gravitational radiation
with a single antenna is very promising, as it would contribute new and very
important data to the understanding of the gravitational interaction, and
also supply experimental evidence for sounder discussions of String Theory.

\newpage

\begin{center} {\large\bf Acknowledgements}
\end{center} We would like to thank V. Fafone and V. Ferrari for useful
discussions. M.B. and M.B. would like to thank C.N.~Colacino for collaboration
at the initial stages of this work. J.A.L. acknowledges financial support from
the Spanish Ministry of Education, contract PB93-1050.

\renewcommand{\theequation}{\thesection.\arabic{equation}}

\appendix

\section{Appendix}

In order to find an analytic expression for $F_n$ defined in (\ref{quarantatre})
we have to perform the following integral
\be \label{appuno} I_n= \int_0^{k_{n2} R} [F_{1(n2)}(r)^2 + 6 F_{2(n2)}(r)^2]
d(k_{n2} r)
\ee where, in accord with (\ref{quarantabis})
\bea  F_{1(n2)}(r)&=&
\b_3(k_{n2}R)k_{n2} r j'_2(q_{n2} r)- 6{q_{n2}\over k_{n2}}\b_1(q_{n2}R)
j_2(k_{n2} r)\nn \\ F_{2(n2)}(r)&=&{k_{n2}\over q_{n2}}\b_3(k_{n2} R)j_2(q_{n2}
r)+ \nn \\ && - {q_{n2}\over k_{n2}}\b_1(q_{n2} R) {d\over d(k_{n2} r)} [k_{n2}r
j_2(k_{n2} r)]
\eea Since 
\be  k^2_{nl}=q^2_{nl}\left(2+{\l\over \mu}\right) 
\ee  choosing $\s_{_P}=1/3$ yields $k_{nl}=2q_{nl}$ and (\ref{appuno}) can be
written  in terms of the following integrals 
\bea  G_1&=&\int_0^{k_{n2}R} j_2(x)^2 dx\nn \\ G_2&=&\int_0^{k_{n2}R} x^2
j_2'(x)^2 dx\nn \\ G_3&=&\int_0^{k_{n2}R} x j_2'(x) j_2(x) dx\nn \\
G_4&=&\int_0^{k_{n2}R} x j_2'(x) j_2(x/2) dx\nn \\ G_5&=&\int_0^{k_{n2}R} x
j_2'(x/2) j_2(x) dx\nn \\ G_6&=&\int_0^{k_{n2}R} j_2(x/2) j_2(x) dx
\eea Because $j_2(x)=(3/x^3-1/x)\sin x-(3/x^2)\cos x$, we simply have to
integrate by parts. For example, let us consider the first integral 
\bea  G_1&=& \int_0^{k_{n2}R} [(9/x^6+1/x^2-6/x^4)\sin^2 x+\nn \\
&+&(9/x^4)\cos^2 x- (6/x^2)(3/x^3-1/x)\sin x\cos x] dx=\nn \\ &=&
\int_0^{k_{n2}R} [9/2x^6+1/2x^2+3/2x^4+\nn \\ &+&(1/2x^2)(15/x^2-9/x^4-1)\cos
2x- (3/x^3)(3/x^2-1)\sin 2x] dx=\nn \\ &=&{Si(2k_{n2}R)\over 5}-{1\over
2(k_{n2}R)}-{1\over 2(k_{n2}R)^3}- {9\over 10(k_{n2}R)^5}+ {\cos(2k_{n2}R)\over
10(k_{n2}R)}+\nn \\ &-&{\sin(2k_{n2}R)\over 5(k_{n2}R)^2}-{13\cos(2k_{n2}R)\over
10(k_{n2}R)^3}+ {9\sin(2k_{n2}R)\over 5(k_{n2}R)^4}+{9\cos(2k_{n2}R)\over
10(k_{n2}R)^5}
\eea where $Si(x)=\int_0^x (\sin x'/x') dx'$. Solving in an analogous way all
the other integrals, we finally obtain 
\bea I_n&=&{1\over 8(k_{n2}R)^5}
\{3\b_1(q_{n2}R)^2 [36\cos(2k_{n2}R) + 72\sin(2k_{n2}R)(k_{n2}R)+ \nn \\ & &-
60\cos(2k_{n2}R)(k_{n2}R)^2 - 24\sin(2k_{n2}R)(k_{n2}R)^3 + \nn \\ & & +
6\cos(2k_{n2}R)(k_{n2}R)^4   + \sin(2k_{n2}R)(k_{n2}R)^5+ \nn \\ & &- 36  -
12(k_{n2}R)^2 - 6(k_{n2}R)^4 + 2(k_{n2}R)^6]+ \nn \\ &+& 16\b_3(k_{n2}R)^2
[1728\cos(k_{n2}R) + 1728\sin(k_{n2}R)(k_{n2}R) +\nn \\ & &-
720\cos(k_{n2}R)(k_{n2}R)^2 - 144\sin(k_{n2}R)(k_{n2}R)^3+ \nn \\ & &+
16\cos(k_{n2}R)(k_{n2}R)^4 +
\sin(k_{n2}R)(k_{n2}R)^5+ \nn \\  & &- 1728 - 144(k_{n2}R)^2  - 16(k_{n2}R)^4 +
(k_{n2}R)^6]+ \nn \\  &-&96\b_3(k_{n2}R)\b_1(q_{n2}R) [36\cos(q_{n2}R) +
18\sin(q_{n2}R)(k_{n2}R) + \nn \\ & &+ 3\cos(q_{n2}R)(k_{n2}R)^2  +
3\sin(q_{n2}R)(k_{n2}R)^3 + \nn \\ & &+ \cos(q_{n2}R)(k_{n2}R)^4  -
36\cos(3q_{n2}R)+ \nn \\ & & - 54\sin(3q_{n2}R)(k_{n2}R)  +
33\cos(3q_{n2}R)(k_{n2}R)^2+ \nn \\  & & + 9\sin(3q_{n2}R)(k_{n2}R)^3 -
\cos(3q_{n2}R)(k_{n2}R)^4]\}
\eea

% \newpage
%%%%%%%%%%%%%%%%%%%%%%%%%%%%%%%%%%%%%%%%%%%%%%%%%%
%%%%%%%%%%%%%%%%%%%%%%%%%%%%%%%%%%%%%%%%%%%%%%%%%%
%%%%%%%%%%%%%%%%%%%%%  BIBLIOGRAFIA   %%%%%%%%%%%%
%%%%%%%%%%%%%%%%%%%%%%%%%%%%%%%%%%%%%%%%%%%%%%%%%%%%%

%%%%%%%%%%%%%%%%%%%%%%%%%%%%%%%%%%%%%%%%%%%%%%%%%%%%%%%%%%%%%%%%%%%%%%%%%%
\end{document}